# Dynamic Model for RNA-seq Data Analysis

Lerong Li[1] and Momiao Xiong[1,*]

[1]Human Genetics Center, Division of Biostatistics, School of Public Health, the University of Texas Health Science Center at Houston, Houston, TX 77030, USA

**Running Title**: RNA-seq data analysis

[*]Address for correspondence and reprints: Dr. Momiao Xiong, Division of Biostatistics, Human Genetics Center, The University of Texas Health Science Center at Houston, 1200 Pressler st, Houston, Texas 77030, (Phone): 713-500-9894, (Fax): 713-500-0900, E-mail: Momiao.Xiong@uth.tmc.edu

**Abstract**

The newly developed deep-sequencing technologies make it possible to acquire both quantitative and qualitative information regarding transcript biology. By measuring messenger RNA levels for all genes in a sample, RNA-seq provides an attractive option to characterize the global changes in transcription. RNA-seq is becoming the widely used platform for gene expression profiling. However, real transcription signals in the RNA-seq data are confounded with measurement and sequencing errors, and other random biological/technical variation. How to appropriately take the variability due to errors and sequencing technology variation into account is essential issue in the RNA-seq data analysis. To extract biologically useful transcription process from the RNA-seq data, we propose to use the second ODE for modeling the RNA-seq data. We use differential principal analysis to develop statistical methods for estimation of location-varying coefficients of the ODE. We validate the accuracy of the ODE model to fit the RNA-seq data by prediction analysis and 5 fold cross validation. We find the accuracy of the second ODE model to predict the gene expression level across the gene is very high and the second ODE model to fit the RNA-seq data very well. To further evaluate the performance of the ODE model for RNA-seq data analysis, we used the location-varying coefficients of the second ODE as features to classify the normal and tumor cells. We demonstrate that even using the ODE model for single gene we can achieve high classification accuracy. We also conduct response analysis to investigate how the transcription process respond to the perturbation of the external signals and identify dozens of genes that are related to cancer.

# 1. Introduction

Next Generation Sequencing (NGS) technologies have revolutionized advances in the study of the transcriptome. The newly developed deep-sequencing technologies make it possible to acquire both quantitative and qualitative information regarding transcript biology. By measuring messenger RNA levels for all genes in a sample, RNA-seq provides an attractive option to characterize the global changes in transcription.

To generate RNA-seq data, the complete set of mRNA are first extracted from an RNA sample and then shattered and reverse transcribed into a library of cDNA fragments with adaptors attached. These short pieces of cDNA are amplified by polymerase chain reaction and sequenced by machine, producing millions of short reads. These reads are then mapped to a reference genome or reference transcript. The number of reads within a region of interest is used as a measure of abundance. The reads can also be assembled de novo without the genomic sequence to create a transcription map.

Compared to microarray which provides limited gene regulation information, RNA-seq offers a comprehensive picture of the transcriptome. RNA-seq has made a number of significant qualitative and quantitative improvements on gene expression analysis and provides multiple layers of resolutions and transcriptome complexity: the expression at exon, SNP, and positional level; splicing; post-transcriptional RNA editing across the entire gene; isoform and allele-specific expression [1].

Many advantages include strong concordance between platforms, higher sensitivity and dynamic range, lower technical variation and background signal, and high level of technical and biological reproducibility, and so on [2-5]. However, some limitations are inherent to next-generation sequencing technology. For example, the read coverage may not be homogeneous along the genome, and different samples may be sequenced at different levels of depth in the experiment. Also, although some genes may have a similar level of expression, longer genes are more likely to have more reads than short ones. Therefore, RNA-seq data must be normalized before any comparison of the counts can be made. Another consideration is that in production of cDNA libraries, larger RNA must be fragmented into smaller pieces to be sequenced and different fragmentations may create bias toward different outcomes. Some other informatics challenges like the storage, transfer, and retrieval of large size data which may bring additional errors [6, 7].

Expression variability measured by RNA-seq arises from three primary sources: (i) real biological differences in different experimental groups or conditions, (ii) measurement errors and (iii) random biological and/or technical variation [1, 8]. The first type of variability is of real biological interest, but is confounded with measurement and sequencing errors, and other random biological/technical variation. How to appropriately take the latter two types of variability into account is essential issue in the RNA-seq data analysis.

Purpose of this paper is to borrow dynamic theory from engineering and use ordinary differential equation (ODE) for modelling the observed number of reads across the gene and unravelling the features of gene transcription [9]. To achieve this goal, we considered the number of reads or expression level at each position as a function of the genomic position and viewed the transcription process as a stochastic process of transcription along the genome. Instead of taking the derivative of expression level with respect to time, we calculated the expression level derivative with respect to genomic position. Specifically, we proposed a dynamic model for the variation of the transcription process along the genome. For each gene, we use a second-order ODE with location-dependent coefficient to model that gene's transcription process. We develop statistical methods for estimation of the coefficient functions in the ODE based on principal differential analysis. Compared to the ODE model with constant coefficient to capture the stochastic variation feature of transcription process, the location-dependent coefficients are essential to account for the complicated stochastic process of gene regulation.

To examine the precision of the ODE for modeling the RNA-seq data, we split the samples into five groups and use 5 fold cross validation to evaluate the accuracy of the predicting gene expression level across the gene using the ODE model.

To capture stochastic feature of gene regulation, we conduct the response analysis. The response analysis of transcriptional processes for each gene using its fitted differential equation can provide important aspects of transcription, including alternative splicing, alternative start and end of transcription, and alternative isoforms. To differentiate feature of gene regulation between normal and cancer tissue samples, we develop statistics to test for significant difference in the response of the gene regulation between the normal and cancer samples under the perturbation of external signals and perform genome-wide response analysis of gene regulation. Using the ODE model, we identified the genes that have a significantly different transcriptional process (both different magnitude and different patterns), and identified genes that showed significantly differential stochastic behaviors in response to environmental perturbations between normal and cancer samples.

To further explore application of the ODE for RNA-seq data analysis, we take the location-varying coefficients of the ODE as features and use FPCA as a tool for extraction of these features. The FPCA scores are used as features and the lasso logistic regression is used as feature selection tool and classifier for distinguishing the cancer and normal samples.

The data suggest that the dynamic features of gene transcription captured by the coefficient functions can retrieve the original process information. Therefore, they are naturally served as a good candidate's features for clustering genes with similar transcription process. These groups of genes could share common biological function, chromosomal location, pathway or regulation. The ODE for modeling the RNA-seq data has the potential to provide valuable information for understanding the mechanism of gene regulation and unraveling disease processes.

## 2. Materials and Methods

*2.1 ODE model with varying coefficients for RNA-seq data.*

Assume that the expression of a gene is measured by the number of sequence reads mapped to this gene in the region $T = [a, b]$. Let $t$ denote a genomic position, $y(t)$ be observed gene expression level that was measured by the number of reads mapped to the genomic position and $x(t)$ be the hidden state that determined the gene expression level at the genomic position $t$. To model transcription process, the second-order ordinary differential equation (ODE) with location varying coefficients can be specified as follow.

$$L(x(t)) = \frac{d^2 x(t)}{dt^2} + w_1(t)\frac{dx(t)}{dt} + w_0(t)x(t) = 0, \qquad (1)$$

where $w_1(t)$ and $w_0(t)$ are weighting coefficients or parameters in the ODE. Its observations $y(t)$ often have measurement errors:

$$y(t) = x(t) + e(t), \qquad (2)$$

where $e(t)$ is measurement error at the position $t$.

*2.2 Estimation of coefficient functions in the ODE*

Estimation of coefficient functions in the ODE consists of two steps. At the first step we estimate the states $\hat{x}(t)$ from the observed number of reads $y(t)$ assuming that coefficient functions in the ODE are given. At the second step, we estimate the coefficient functions in the ODE, assuming that states $x(t)$ have been estimated.

**Step 1:**

To estimate $x(t)$, we first expand the function $x(t)$ in terms of basis functions $\phi(t)$ and then estimate its expansion coefficients. Let $x_i(t)$ be the state variable at the genomic position $t$ of the $i$-th sample and $y_i(t)$ be its observation ($i = 1, \ldots, n$). Then, $x_i(t)$ can be expanded as

$$x_i(t) = \sum_{j=1}^{K} c_{ij}\phi_j(t) = C_i^T \phi(t) \tag{3}$$

Where $C_i = [c_{i1}, \ldots, c_{iK}]^T$ and $\phi(t) = [\phi_1(t), \ldots, \phi_K(t)]^T$

Similarly, the parameters $w_1(t)$ and $w_0(t)$ can be expanded as

$$w_1(t) = \sum_{j=1}^{K} h_{1j}\phi_j(t) = h_1^T \phi(t) \text{ and}$$

$$w_0(t) = \sum_{j=1}^{K} h_{0j}\phi_j(t) = h_0^T \phi(t). \tag{4}$$

Substituting their expansions into equation (1), we obtain

$$L(x_i(t)) = C_i^T \Psi(t), \tag{5}$$

where $\Psi(t) = \dfrac{d^2\phi}{dt^2} + G(t)h$, $G(t) = [\dfrac{d\phi}{dt}\phi^T(t), \phi(t)\phi^T(t)]$, $h = [h_1^T, h_0^T]^T$. To smooth the estimated function $\hat{x}(t)$, we impose the following penalty term:

$$\lambda \int_T L(x_i(t))L^T(x_i(t))dt = \lambda C_i^T J_{\phi h} C_i,$$

where $J_{\phi h} = \int_T \Psi(t)\Psi^T(t)dt$.

We estimate the state function $x(t)$ from the observation data $y(t)$ by minimizing the following objective function which consists of the sum of the squared errors between the observations and the estimated states, and the penalty terms:

$$\sum_{i=1}^{n}\{\sum_{j=1}^{\tau}[y_i(t_j) - x_i(t_j)]^2 + \lambda \int_T L(x_i(t))L^T(x_i(t))dt\}$$
$$= \sum_{i=1}^{n}\{\sum_{j=1}^{\tau}[y_i(t_j) - x_i(t_j)] + \lambda C_i^T J_{\phi h} C_i\} \quad (6)$$

Let

$Y_i = [y_i(t_1),..., y_i(t_T)]^T$, $Y = [Y_1^T,..., Y_n^T]$, $\tilde{\phi} = [\phi(t_1),..., \phi(t_T)]^T$, $C = [C_1^T,..., C_n^T]^T$,

$\Phi = daig(\tilde{\phi},...,\tilde{\phi})$ and $J = diag(J_{\phi h},..., J_{\phi h})$.

Problem (6) can be rewritten in a matrix form:

$$\min_C (Y - \Phi C)^T(Y - \Phi C) + \lambda C^T J C$$

The least square estimators of the expansion coefficients are then given by

$$C = (\Phi^T \Phi + \lambda J)^{-1} \Phi. \quad (7)$$

**Step 2**

Next we estimate the coefficient functions in the ODE. The coefficient functions in the ODE can be estimated by minimizing the following least squares objective function:

$$\min_h SSE_p = \int_T L^T(\hat{X}(t))L(\hat{X}(t))dt, \quad (8)$$

where $L(\hat{X}(t)) = [L(\hat{x}_1(t)),..., L(\hat{x}_n(t))]^T$.

Since $L(x_i(t)) = C_i^T \Psi(t)$, the $L(x(t))$ can be expressed in terms of the estimated expansion coefficients as

$$L(\hat{X}(t)) = C_* \Psi(t),$$

where the matrix $C_*$ is defined as $C_* = \begin{bmatrix} C_1^T \\ \vdots \\ C_n^T \end{bmatrix}$

Therefore, problem (8) can be reduced as

$$\min_{h} \ SSE_p = \int_T \Psi^T(t) C_*^T C_* \Psi(t) dt. \tag{9}$$

where the matrix $C_*$ is estimated and hence fixed in the minimization problem (9). Setting the partial derivative of $SSE_p$ to be zero:

$$\frac{\partial SSE_p}{\partial h} = \int_T G^T(t) C_*^T C_* \left[ \frac{d^2 \phi}{dt^2} + G(t) h \right] dt = 0. \tag{10}$$

Solving equation (10) for $h$, we obtain

$$h = -\left[ \int_T G^T(t) C_*^T C_* G(t) dt \right]^{-1} \int_T G^T(t) C_*^T C_* \frac{d^2 \phi}{dt^2} dt. \tag{11}$$

In summary, we iteratively determine the expansion coefficients of the state function $X(t)$ for fixed parameters in the ODE by equation (7) and estimate the coefficient functions in the ODE for fixed expansion coefficients by equation (11).

*2.3. ODE for classification*

To illustrate that the ODE can be used as a useful tool for modeling the profile of the RNA-seq expression we will show that the ODE can capture all variation of gene expression across the gene and that the coefficient functions of the ODE are useful feature extraction of the RNA-seq data. The ODE can be used for classifying tumor and normal samples.

Since dimensions of the coefficient functions of the ODE are extremely high, the functional principal component analysis (FPCA) is used to reduce the dimensions of the coefficient functions of the ODE.

The FPCA tries to find the dominant direction of variation around an overall trend function [10, 11] Each principal component is specified by the weight function $\beta(t)$, and the principal component scores of the individuals in the sample are defined as the inner product of weight function and functional curves ($w_0(t), w_1(t)$).

$$z = \int_T \beta(t)w(t)dt ,$$

where for convenience, we use $w(t)$ to denote either $w_1(t)$ or $w_0(t)$. That is the coordinate value of functional curves at the direction of $\beta(t)$ with highest variability. By projecting the functional curves onto set of eigenfunctions, we can reduce the dimension to finite number, functional principal component scores.

Suppose that for the $i$-th individual sample we obtain the functional principal component score:

$$z_{ij}^{(1)} = \int_T \beta_j(t)w_{i1}(t)dt \text{ and } z_{ij}^{(0)} = \int_T \beta_j(t)w_{i0}(t)dt ,$$

where $w_{i1}(t)$ and $w_{i0}(t)$ are the coefficient functions of the ODE for he $i$-th individual sample, and $\beta_j(t), j = 1,...,K$ are a set of eigenfunctions (or principal component functions). The original functional curves can be reduced to a finite feature matrix:

$$Z = \begin{bmatrix} z_{11}^{(0)} & z_{11}^{(1)} & \cdots & z_{1K}^{(0)} & z_{1K}^{(1)} \\ \vdots & \vdots & \ddots & \vdots & \vdots \\ z_{n1}^{(0)} & z_{n1}^{(1)} & \cdots & z_{nK}^{(0)} & z_{nK}^{(1)} \end{bmatrix}$$

where the $K$ is the number of principal components selected to explain the total variability.

To improve classification accuracy we use the LASSO logistic regression as a classifier. In simple logistic regression, we use the logit link to relate the mean of response with the covariates of interest. Let $x_i = [x_1, ..., X_p]^T$ be the vector of observed covariates for ith observation, and $y_i$ is the corresponding response outcome. For simplicity, we consider binary cases where $y_i = 1\ or\ 0$. The model is specified as the following posterior probability for $i$th observation[12].

$$\pi_i(x_i, \beta) = \Pr(y_i = 1 | x_i, \beta) = \frac{\exp(\beta_0 + \beta^T x_i)}{1 + \exp(\beta_0 + \beta^T x_i)}$$

where $\beta = [\beta_1, ..., \beta_p]^T$ is the covariate vector of interest, and $\beta_0$ is the intercept term. And the joint log-likelihood of the N subjects is defined as

$$l(\beta) = \sum_{i=1}^{N} \log \pi_i(x_i, \beta)$$

which can be written as

$$l(\beta) = \sum_{i=1}^{N} \{y_i \log \pi_i(x_i, \beta) + (1-y_i)\log(1 - \pi_i(x_i, \beta))\}$$

$$= \sum_{i=1}^{N} \{y_i(\beta_0 + \beta^T x_i) - \log(1 + e^{\beta_0 + \beta^T x_i})\}$$

To estimate the parameter, we set its derivatives to zero and get the score equations

$$\frac{\partial l(\beta)}{\partial \beta} = \sum_{i=1}^{N} x_i(y_i - \pi_i(x_i, \beta)) = 0 \quad (12)$$

Since (12) is nonlinear equations in $\beta$, we usually use some iterative methods like Newton-Raphson algorithm to get the solution of $\beta$.

By adding a $L_1$ penalty to the joint log-likelihood in equation (12) we have the following constrained maximization equation

$$l_c(\beta) = \left\{ \sum_{i=1}^{N} \{y_i(\beta_0 + \beta^T x_i) - \log(1 + e^{\beta_0 + \beta^T x_i})\} - \lambda \sum_{j=1}^{p} |\beta_j| \right\} \quad (13)$$

$l_c(\beta)$ is the constrained log-likelihood and $\lambda$ is tuning parameter to adjust the tradeoff between log-likelihood function and the size of penalty. Please note that in Lasso, we usually do not penalize the intercept term and it is practical meaningful to standardize the covariates before optimization.

The $L_1$ penalty is not differentiable and also $\beta$ is not linear solution of response **y.** It is not trivial to get the score functions but we can still have a solution using nonlinear programming method [13]. The score functions for variables with non-zero coefficients have the form

$$x_j^T(y - \pi) = \lambda \cdot sign(\beta_j)$$

Coordinate descent method is one efficient method to compute the Lasso solution. It fixes the penalty parameters $\lambda$ and optimize over each parameter successively, while holding the others fixed at current values. R package *glmnet* [14] can efficiently fit the Lasso logistic regression with large $N$ and $p$.

*2.4 Numerical solution to the ODE with bounded values.* We use collocation Runge-Kutta method for the solution of boundary value problem of the ODE. The basic idea is to find a set of polynomials $p_n(x)$ of degree s which satisfies the problem over the interval $[x_{n-1}, x_n]$ for a set of points

$$x_{nj} = x_{n-1} + a_j h_n, \text{ where } j = 1,...,s \text{ and } n = 1,...,N$$

Note that $0 < a_0 < a_1 < ... < a_s < 1$, they are distinct real numbers. Also the polynomial functions $p_n(x)$ are set to satisfy

$$\begin{aligned} p_n(x_{n-1}) &= y_{n-1} \\ p_n'(x_{nj}) &= f(x_{nj}, p(x_{nj})) \end{aligned}, \text{ where } j = 1,...,s$$

The numerical approximation at $x_n$ is given by

$$y_n = p_n(x_{n-1} + h_n)$$

R package **bvpSolve** implements the method for boundary value problem [15].

*2.5 Response analysis under perturbation of external signal.*

Gene regulatory properties are encoded in the parameter curves of the ODE modeling gene expressions. Testing significant difference in the parameter curves between two conditions can be used as a powerful tool to assess differential changing behaviors of the gene expression across the gene region between two conditions. Response analysis attempts to extract inherent features of the systems that capture and describe the behaviors of the system over genomic positions under different operating conditions and perturbation of external signals.

Let $t$ denote a genomic region within the gene of interests and $x(t)$ be the number of reads mapped to the genomic region. And the ODE model used to describe the expression profile is given as follow

$$L(x(t)) = \frac{d^2 x(t)}{dt^2} + w_1(t)\frac{dx(t)}{dt} + w_0(t)x(t) = 0$$

Suppose the $\widehat{w_1}(t)$ and $\widehat{w_0}(t)$ are estimated from the data. The response of a regulatory system depends on the input signals. Different signal will cause different responses. For simplicity, we consider unit-step signal forced on the system and then solve the responses of the original system between different groups using estimated parameters $\widehat{w_1}(t)$ and $\widehat{w_0}(t)$.

$$\frac{d^2 x(t)}{dt^2} + \widehat{w_1}(t)\frac{dx(t)}{dt} + \widehat{w_0}(t)x(t) = U(t)$$

To solve the solution of the estimated ODE with unit-step force function $U(t)$, we have to use some numerical methods to approximate the solution $\hat{x}(t)$. We solved ODE numerically by considering two-point boundary value problems where boundary conditions are specified at both ends of the range of integration. We estimated two initial values at both ends by evaluating the estimated smoothing expression curves at start and end positions.

Suppose $R(t) = [r_1(t), r_2(t), \ldots, r_{N_1}(t)]^T$ be a vector-valued function to represent response functional for all $N_1$ subjects in the normal group and $S(t) = [s_1(t), s_2(t), \ldots, s_{N_2}(t)]^T$ be response functional for $N_2$ subjects in cancer group. Therefore, we can construct a Hotelling $T^2$. Suppose that the response functions were expanded in terms of eigenfunctions $\phi_1(t), \ldots, \phi_K(t)$:

$$r_i(t) = \sum_{j=1}^{K} \xi_{ij} \phi_j(t),$$

$$s_i(t) = \sum_{j=1}^{K} \eta_{ij} \phi_j(t),$$

Where $\xi_{ij} = \int_T r_i(t)\phi_j(t)dt$ and $\eta_{ij} = \int_T s_i(t)\phi_j(t)dt$, $\xi_{ij}$ and $\eta_{ij}$ are uncorrelated random variables with zero mean and variances $\lambda_j$ with $\sum_j \lambda_j < \infty$. Define the averages $\bar{\xi}_j$ and $\bar{\eta}_j$ of the principal component scores $\xi_{ij}$ and $\eta_{ij}$ in the normal and cancer group. Then we denote the average vector of scores in normal and cancer group by

$$\bar{\xi} = [\bar{\xi}_1, \ldots \bar{\xi}_k]^T$$

$$\bar{\eta} = [\bar{\eta}_1, \ldots \bar{\eta}_k]^T$$

where $\bar{\xi}_j = \frac{1}{N_1}\sum_{i=1}^{N_1} \xi_{ij}$ and $\bar{\eta}_j = \frac{1}{N_2}\sum_{i=1}^{N_2} \eta_{ij}, j = 1, \ldots, k$

The pooled covariance matrix is

$$S = \frac{1}{N_1 + N_2 - 2}\left(\sum_{i=1}^{N_1}(\xi_i - \bar{\xi})(\xi_i - \bar{\xi})^T + \sum_{i=1}^{N_2}(\eta_i - \bar{\eta})(\eta_i - \bar{\eta})^T\right)$$

where $\xi_i = [\xi_{i1}, \ldots, \xi_{ik}]^T$ and $\eta_i = [\eta_{i1}, \ldots, \eta_{ik}]^T$

Let $\Lambda = (\frac{1}{N_1} + \frac{1}{N_2})S$, then the Hotelling statistics can be written as

$$T^2 = (\bar{\xi} - \bar{\eta})\Lambda^{-1} = (\bar{\xi} - \bar{\eta})^T$$

Under null of no difference in the response of the gene regulation between two groups, the statistics follows $\chi_k^2$ distribution where $k$ is the number of principle component scores.

## 3. Results

*3.1 Dataset.* We apply the proposed model to Kidney Renal Clear Cell Carcinoma (KIRC) RNA-seq data, which is available from The Cancer Genome Atlas (TCGA) project (https://tcga-data.nci.nih.gov/tcga/). The RNA-seq data is available for 72 matched pair of KIRC and normal samples. The maximum number of genomic positions where the expressions were measured by the number of reads passing qualify of control is 382,239,893 in the raw BAM file. And the total number of genes is 19,717.

Samtools and bedtools were applied to count number of reads for each base of the gene. Effected mapping reads was taken as the scale factor to normalize the reads for each individual. Hg19 human genome was taken as the reference.

Illumina paired-end RNA sequencing reads were aligned to GRCh37-lite genome-plus-junctions reference using BWA version 0.5.7. This reference combined genomic sequences in the GRCh37-lite assembly and exon-exon junction sequences whose corresponding coordinates were defined based on annotations of any transcripts in Ensembl (v59), Refseq and known genes from the UCSC genome browser, which was downloaded on August 19 2010, August 8 2010, and August 19 2010, respectively. Reads that mapped to junction regions were then repositioned back to the genome, and were marked with 'ZJ:Z' tags. BWA is run using default parameters, except that the option (-s) is included to disable Smith-Waterman alignment. Finally, reads failing the Illumina chastity filter were flagged with a custom script, and duplicated reads were flagged with Picard's MarkDuplicates.

In order to make the data comparable, we applied log transformation on the observed expression profiles. Some genomic position has zero counts and we intentionally add 1 to it and then it returns to be zero after log transformation. After that expression counts for most of genes are of the same scale. We also mapped the genes onto the interval [0,100].

*3.2. Evaluation of the ODE for modeling RNA-seq data.*

To evaluate the precision of the ODE for modeling the RNA-seq data, we first used the ODE to fit the RNA-seq data where the coefficient functions were estimated. Then, we used numerical collocation Runge-Kutta method to solve the fitted ODE. The solutions of the fitted ODE as a function of genomic position were then compared with the observed RNA-seq curves.

We estimated the varying-coefficient functions using the proposed model. The expression function for gene $X(t)$ was first estimated by spline smoothing with some initial penalty. We then update the penalty using the proposed second order ODE with varying-coefficient functions. We iterated between curve smoothing and ODE estimation until convergence was achieved. The smoothing parameters $\lambda$ were chosen by cross-validation process. By selecting the value of $\lambda$, we trade off basis expansion fitting error and ODE solution filtering error. Larger value of $\lambda$ put more emphasis on the ODE penalty and the solution to ODE with estimated parameters are more likely to approximate the original data.

To validate the estimates of coefficients functions in the model, it is essential to compare the observed gene expression curve to the ODE solution with estimated coefficient functions. We solved ODE numerically by considering two-point boundary value problems where boundary conditions are specified at both ends of the range of integration. We estimated two initial values at both ends by evaluating the estimated smoothing expression curves at start and end positions.

Figures 1A and 1B are fitted results of normal sample and cancer sample respectively for gene CD74. In these figures, the circles represent observed RNA-seq expression signal (green: normal; red: cancer); the blues lines are Fourier basis expansion to approximate the observed signal using weighted least square methods. The numbers of basis are chosen based on the length of genes and experimental adjustment to capture the important characteristics of gene expression curves. The dashed lines are estimated ODE solution using boundary value problem solver (R package **bvpSolve).** The ODE solutions approximate well the observed expression level of gene *CD74*, which show that estimated coefficient functions carry essential information of the original data. Once we have them, we can retrieve the original data very well.

To further evaluate the precision of the ODE for modeling RNA-seq data, we perform 5-fold cross-validation prediction for gene *RPL29*. This method uses part of the available data to fit the model and estimate the parameters, and uses the remaining data to test the model validity and estimate accuracy. We randomly split normal and cancer samples into 5 folds. From the estimation of parameters in the training samples, we solved the ODE with estimated coefficient functions to predict the expression curves of test samples. To be consistent, we estimated two initial values at both ends by evaluating the estimated smoothing expression curves at start and end positions in test samples. We also calculated the root mean square prediction error (RMSPE) for each folder to evaluate the performance of the prediction which is defined by

$$RMSPE_j = \frac{1}{N_j} \sum_{j=1}^{N_j} \sqrt{\frac{1}{N} \sum_{i=1}^{N} (\hat{y}_i - y_i)^2} \qquad (4.1)$$

Where $y_i$ and $\hat{y}_i$ are observed and predicted expression level; $N$ is the number of genomic positions where the RNA-seq are observed for the gene and $N_j$ is the number of subjects in the folder j.

The table 1 lists RMSPE in each folder for normal and cancer groups. The normal group has slightly better performance in terms of prediction on the test samples. But both prediction errors are relatively small.

Figures 2A and 2B are prediction results for selected samples in test set for gene RPL29 in normal and cancer group respectively. The gray dot is observed expression profile, and the solid red lines are Fourier basis expansion approximations to the observed expression data. The dashed green lines indicate the predicted gene expression profile in the test set by solving the estimated ODE in the training examples. We can observe that all of the prediction can capture the overall shape and fluctuation in the data. Secondly, they can also predict the magnitude of expression value with decent accuracy. These are predicted very close to the observed expression profiles.

*3.3 Classification analysis.*

These data suggest that the estimated coefficient functions capture important features of expression curves. From the solution to estimated ODE, we can see the exceptional retrieval of original data. From the prediction performance in the test set, we can also get well predicted curves by just proving two initial boundary data points. It is natural to consider them as features to classify phenotype categories.

We obtain two coefficient functions from one expression function. We can use FPCA to help us to reduce the dimension of features and to ease the computational effort. We first applied FPCA technique on two coefficient functions $w_0(t)$ and $w_1(t)$ separately, then we combined two groups of the selected functional principal component scores as aggregated features before we provided them to classifier. In the end, we applied lasso logistic regression to help us select features and make prediction on the groups.

The table 2 lists top 12 genes to differentiate normal and KIRC group using 5 fold cross-validations. We can see that using a single gene it can reach as high as 90% classification accuracy. These data strongly indicate that the ODE model effectively captured the inherent features of RNA-seq expression profile. We also evaluated the performance of classification result using sensitivity, specificity and accuracy. Sensitivity is defined as the percentage of cancer tissues correctly classified as cancer. Specificity is defined as the percentage of the normal tissues correctly classified as normal. The classification accuracy is defined as the percentage of the correctly classified normal and cancer tissues. The classification results can be reached as high as 99% if we use all these 12 genes together as predictors.

*3.4 Genome-scale clustering analysis.*

In this section, we continue to use the estimated coefficient functions as features to cluster genes expression data to study the genome wide transcriptome. By grouping genes with similar patterns of expression profiles, cluster analysis can provide insight into gene functions and biological process. It also gives a simple way of determining the functions of many genes for which information is not available, as genes with same functions may share expression profiles. We assume the coefficient functions in ODE model help to define these patterns in the dynamic regulation process and give us clues to functional discovery and pattern grouping.

After we derive the feature matrix for all the genes from dimension reduction using FPCA, we merely need to adopt a metric definition which is used as a measure of similarity in the behavior of two genes. To calculate the distance matrix we used Euclidean distance and correlation matrix. This method computes a dendrogram that combine all genes in a single tree.

A total of 19717 genes were clustered into 9 groups according to the cluster analysis (Figures 3A and 3B). The functional principal component scores from coefficient functions in ODE model were used as significant features to define these patterns in the dynamic regulation process. The function annotation for each cluster was as the following.

The principle functions of the genes in the first group are mainly associated with oxidoreductase activity, ligase activity, dehydrogenase (NAD) activity and related metabolic process. The detail functions include aldehyde dehydrogenase (NAD) activity, translational initiation, mediator complex, MHC protein complex, mitochondrial membrane part, ion transmembrane transport, respiratory chain complex I, proton-transporting ATP synthase complex, proton-transporting two-sector ATPase complex, proton-transporting domain, NADH dehydrogenase (quinone) activity, oxidoreductase activity, acting on the aldehyde or oxo group of donors, NAD or NADP as acceptor, positive regulation of protein ubiquitination, response to unfolded protein, heterocycle metabolic process, protein modification process, mitochondrial ATP synthesis coupled proton transport, glycerolipid metabolic process, macromolecule modification, proton-transporting two-sector ATPase complex, catalytic domain, regulation of translational initiation, oxidoreductase activity, acting on the aldehyde or oxo group of donors, RNA polymerase II transcription mediator activity, heme binding, positive regulation of ligase activity, negative regulation of ligase activity, negative regulation of ubiquitin-protein ligase activity involved in mitotic cell cycle, positive regulation of ubiquitin-protein ligase activity involved in mitotic cell cycle, regulation of ubiquitin-protein ligase activity involved in mitotic cell cycle, positive regulation of ubiquitin-protein ligase activity, negative regulation of ubiquitin-protein ligase activity, transferase activity, glycerolipid biosynthetic process, amine transport, phosphoinositide metabolic process, carboxylic acid transport, hormone binding, eukaryotic translation initiation factor 3 complex, glycerophospholipid metabolic process, helicase activity, response to protein stimulus, lipid biosynthetic process, phosphoinositide biosynthetic process, aldehyde dehydrogenase [NAD(P)+] activity, proton-transporting ATP synthase complex, coupling factor F(o), cytosolic part, nucleobase, nucleoside and nucleotide metabolic process, proton-transporting ATPase activity, rotational mechanism, phospholipid metabolic process, phosphorus metabolic process, phosphate metabolic process, hydrogen-exporting ATPase activity, phosphorylative mechanism, proton-transporting V-type ATPase complex, MHC class II protein complex, collagen, positive regulation of protein modification process and post-translational protein modification.

The principle functions of the genes in the second group are mainly associated with hydratase activity, cation transmembrane transporter activity and hydrolase activity and related metabolic process. The detail functions include NAD or NADH binding, peroxisomal membrane, microbody membrane, aconitate hydratase activity, 4 iron, 4 sulfur cluster binding, regulation of vesicle-mediated transport, lactate dehydrogenase activity, L-lactate dehydrogenase activity, long-chain fatty acid-CoA ligase activity, fatty acid ligase activity, homophilic cell adhesion, tight junction, cation transmembrane transporter activity, occluding junction, kinesin complex, microbody part, peroxisomal part, actin filament binding, hydrolase activity, hydrolyzing O-glycosyl compounds, hydrolase activity and acting on glycosyl bonds.

The principle functions of the genes in the third group are mainly associated with monooxygenase activity, receptor activity, electron carrier activity, sodium ion transmembrane transporter activity and related metabolic process The detail functions include nucleoside binding, purine nucleoside binding, monooxygenase activity, receptor activity, protein binding, ATP binding, DNA packaging, chromatin assembly or disassembly, electron carrier activity, sodium ion transmembrane transporter activity, actin cytoskeleton, RNA metabolic process, adenyl nucleotide binding, GTPase regulator activity, regulation of lipid transport, negative regulation of lipid transport, adenyl ribonucleotide binding, protein-DNA complex, very-low-density lipoprotein particle, triglyceride-rich lipoprotein particle, cellular nitrogen compound metabolic process, cellular macromolecule biosynthetic process, nucleosome organization, chylomicron, organelle, intracellular organelle, cellular biosynthetic process, keratin filament, regulation of transcription, regulation of biological process, regulation of cellular process, regulation of nitrogen compound metabolic process, regulation of RNA metabolic process, regulation of macromolecule metabolic process, nucleoside-triphosphatase regulator activity, biological regulation, DNA conformation change and regulation of primary metabolic process.

The principle functions of the genes in the fourth group are mainly associated with acyl-CoA thioesterase activity, oxidoreductase activity, phosphatase activity and related metabolic process. The detail functions include organellar small ribosomal subunit, organellar large ribosomal subunit, phospholipid-translocating ATPase activity, glutathione transferase activity, receptor signaling protein serine/threonine kinase activity, transmembrane receptor activity, inward rectifier potassium channel activity, organic acid transmembrane transporter activity, mitochondrial matrix, mitochondrial large ribosomal subunit, mitochondrial small ribosomal subunit, cytosol, translation, translational elongation, cell surface receptor linked signaling pathway, large ribosomal subunit, small ribosomal subunit, integral to membrane, acyl-CoA thioesterase activity, oxidoreductase activity, acting on NADH or NADPH, phosphatase activity, cytosolic ribosome, signaling process, signal transmission, intrinsic to membrane, negative regulation of protein ubiquitination, cullin-RING ubiquitin ligase complex and mitochondrial lumen.

The principle functions of the genes in the fifth group are mainly associated with cell projection part, microtubule associated complex, motor activity, microtubule, axoneme, microtubule-based process, microtubule-based movement, microtubule cytoskeleton, dynein complex, cytoskeletal part, cilium, macromolecular complex, cilium axoneme, cell projection, protein complex, cilium part, pyrophosphatase activity, hydrolase activity, acting on acid anhydrides, hydrolase activity, acting on acid anhydrides, in phosphorus-containing anhydrides, nucleoside-triphosphatase activity.

The principle functions of the genes in the sixth group are mainly associated with intracellular signal transduction, cholesterol efflux, UDP-galactosyltransferase activity and histone demethylase activity.

The principle functions of the genes in the seventh group are mainly associated with ATP-binding cassette (ABC) transporter complex, JNK cascade, ATP-dependent peptidase activity,

The principle functions of the genes in the eighth group are mainly associated with glutamate receptor activity, ATPase activity, coupled, cytoskeletal protein binding, myosin filament,

The principle functions of the genes in the ninth group are mainly associated with adrenoceptor activity, inhibition of adenylate cyclase activity by G-protein signaling pathway, adenosine deaminase activity, hydrolase activity, acting on carbon-nitrogen (but not peptide) bonds, in cyclic amidines, deaminase activity, adenylate cyclase activity, activation of protein kinase A activity, alpha-adrenergic receptor activity, adrenergic receptor binding, epinephrine binding, regulation of norepinephrine secretion, norepinephrine transport, positive regulation of blood pressure, norepinephrine secretion, oxidoreductase activity, acting on CH-OH group of donors, oxidoreductase activity, acting on the CH-OH group of donors, NAD or NADP as acceptor and delayed rectifier potassium channel activity.

*3.5 Response analysis of gene regulation.*

The expression level of a gene measured by sequencing can be viewed as a curve or function of genomic position. The gene expression will vary across the gene region. If we treat time and space position as the same argument, all theory and methods of dynamic system can be applied to RNA-seq data analysis. The dynamic behavior of a system is encoded in the temporal evolution of its states or in the genomic location evolution of the gene expression in our problem. Therefore, borrowing dynamic theory, we can study the location-dependent variation of gene expression under the perturbation of the external signals. The transient response of the dynamic systems is an important property of the system itself. It can be used to quantify the space domain characteristics of the gene regulation system responding to the disturbance of environments. Our goal is to investigate how the gene expression level at each genomic position varies in response to the external perturbation and whether this will affect the function of cell.

We conducted response analysis of 19,717 genes under unit-step signal perturbation. We used the Hoteling $T^2$ statistic that was described in Section 2.5 to identify 31 genes that showed significant difference in the response property. The names of 31 genes with significant difference in response property were summarized in Table 3.

We present Figures 4A-D showing the average expression curves, unit-step response curves, the coefficient curves of the ODE of gene CD74, respectively. We observed that gene CD74 not only showed significant difference in gene expression and coefficient curves of the ODE, but also demonstrated strong difference in the unit-step response. The changing point of gene expression curve and unit-step response curve occurred between 11b and 12 a where a splicing site is located. It was reported that CD 74 played critical role in cancer cell tumorigenesis [16] and down-regulation of CD74 inhibits growth and invasion in clear cell renal cell carcinoma [17].

Transient response is one of dynamic property. As Figures S1A-D shown, gene *ABHD10* that did not show significant difference in gene expression and coefficient curves of the ODE, but demonstrated strong difference in the unit-step response.

Figures 2A-D plotted the average expression curves, unit-step response curves, the coefficient curves of the ODE of gene BTS2, respectively. Gene *BTS2* was differentially expressed, but did not show significant difference in coefficient of the ODE between tumor and normal samples. Gene *BTS2* was identified to have significant difference in the unit-step response. The pattern of difference in the unit-step response may mainly due to rapid changes of gene expression in the region close to genomic position 20. From the literature we found that *BTS2* was associated with a number of cancers [18, 19].

## 4. Discussion

Dominant methods in literature for RNA-seq data analysis use a single valued summary statistic to represent expression level of a gene. However, a single number oversimplifies complex expression variation pattern across a gene and ignore information on alternative splicing, isoform and expression level variation at the genomic position level. To extract biologically useful expression variation signals across gene from RNA-seq data which are confounded with the sequencing technology variation is a challenge, but important task. To meet this challenge, we have proposed to use the ODE for modeling the RNA-seq data and addressed several essential issues for application of the ODE model to RNA-seq data analysis.

The first issue is how to use the ODE for modeling the RNA-seq data. We considered the number of reads or expression level at each position as a function of the genomic position and viewed the transcription process as a stochastic process of transcription along the gene. Borrowing dynamic theory from engineering, we have used the second ODE to model the expression function of the gene measured by RNA-seq. We have employed differential principal analysis to develop statistical methods for estimation of location-varying coefficients of the ODE.

The second issue is the precision of the ODE to model the RNA-seq data. We randomly split normal and cancer samples into 5 folds. From the estimation of parameters in the training samples, we solved the ODE with estimated coefficient functions to predict the expression curves of test samples. We have showed that the accuracy of the prediction by the second ODE was very high and the root mean square prediction errors were quite small.

The third issue is how to extract useful regulatory signals from the RNA-seq data confound with measurement errors and sequencing technology variation. Since the second ODE can model RNA-seq data very well, the location-coefficient functions of the ODE may well characterize the features of the regulatory process and measure the impact of the gene expression on the function of the cells and tissues. We have demonstrated that using location-coefficient functions of the second ODE as features we have accurately classify the tumor and normal samples.

The fourth issue is to explore the applications of the ODE for RNA-seq data analysis. We have showed that the ODE can be used as a powerful tool to study the response of the gene transcription to the perturbation of environments. We have identified a number of cancer associated genes which showed significant difference in the response of the gene transcription between tumor and normal tissues, but were not differentially expressed.

To our knowledge, this is the first to use the ODE for modeling the RNA-seq data and investigation of gene transcription process. Our results were very preliminary. The samples were used to validate the accuracy of the ODE model to fit the real RNA-seq data. Large-scale validation and experiments for evaluating the model precision are urgently needed. Although the response analysis of dynamic model for the transcription process can help us to study how the external signals affect the gene expression variation across the gene, the mechanism of the gene transcription variation under the perturbation of external signals are largely unknown. The experiments for validation the results of the response analysis of the dynamic models need to be performed. We are lack of consensus methods for RNA-seq data analysis. We are facing great challenges in developing innovative approaches and general framework for RNA-seq data analysis.

## 5. Conclusions

In conclusion, this study propose the second ODE for modeling RNA-seq data. We have demonstrated that the estimated ODE can accurately predict the gene expression level across the gene. We have showed that the location-dependent coefficients of the ODE effectively extract regulatory signals from the RNA-seq confounded with the measurement errors and sequencing technology variation and capture the inherent features of the transcription process. The results have showed that using coefficients of the ODE as features we can reach very high accuracy for classifying tumor and normal samples. Finally, we have demonstrated that using transient response analysis of dynamic system we identify 31 genes with significant differential response behavior between tumor and normal samples are related to cancer.

**Acknowledgments**

The project described was supported by Grant 1R01AR057120–01 and 1R01HL106034-01 from the National Institutes of Health and NHLBI. The authors wish to thank for providing RNA-seq data. The authors wish to acknowledge the contributions of the research institutions, study investigators, field staff and study participants in creating the TCGA datasets for biomedical research.

**Conflict of interest**

The authors declare that there is no conflict of interests regarding the publication of the paper.

Table 1. RMSPE in each folder for normal and cancer groups

| Folder list | Normal RMSPE | Cancer RMSPE |
|---|---|---|
| 1 | 0.23 | 0.97 |
| 2 | 0.31 | 0.94 |
| 3 | 0.24 | 0.79 |
| 4 | 0.33 | 0.70 |
| 5 | 0.30 | 0.73 |

Table 1. The average sensitivity, specificity and accuracy of top 12 genes to classify normal and KIRC group over 5-fold cross validation.

| Genes | Sensitivity | Specificity | Accuracy |
|---:|:---:|:---:|:---:|
| *RBBP8* | 0.903 | 0.958 | 0.931 |
| *ZFYVE16* | 0.903 | 0.958 | 0.931 |
| *LOC100129034* | 0.889 | 0.944 | 0.917 |
| *SLC44A2* | 0.931 | 0.903 | 0.917 |
| *TTC21B* | 0.903 | 0.931 | 0.917 |
| *C18orf56* | 0.958 | 0.861 | 0.910 |
| *KCNJ16* | 0.889 | 0.931 | 0.910 |
| *PFKP* | 0.917 | 0.903 | 0.910 |
| *TMCC1* | 0.903 | 0.903 | 0.903 |
| *CDK18* | 0.917 | 0.875 | 0.896 |
| *SEC61G* | 0.903 | 0.889 | 0.896 |
| *ST6GAL1* | 0.861 | 0.931 | 0.896 |

Table 3. Genes with significantly difference in response behavior between normal and tumor samples.

| | | | | | |
|---|---|---|---|---|---|
| *ABHD10* | *MFSD1* | *SDR39U1* | *ATP6V1D* | *OXA1L* | *SEC31A* |
| *BST2* | *PACSIN2* | *SMCR8* | *CD74* | *PGAM1P5* | *SSR2* |
| *DAP3* | *PIK3CB* | *TM9SF2* | *DHX40* | *PITRM1* | *UBXN6* |
| *EDF1* | *POLR2B* | *UQCRC2* | *HLA-DMB* | *PSAP* | *VKORC1* |
| *HSPA9* | *PSMB10* | *ZNF710* | *ISYNA1* | *MAT2A* | *PSMB7* |
| *PSMC4* | | | | | |

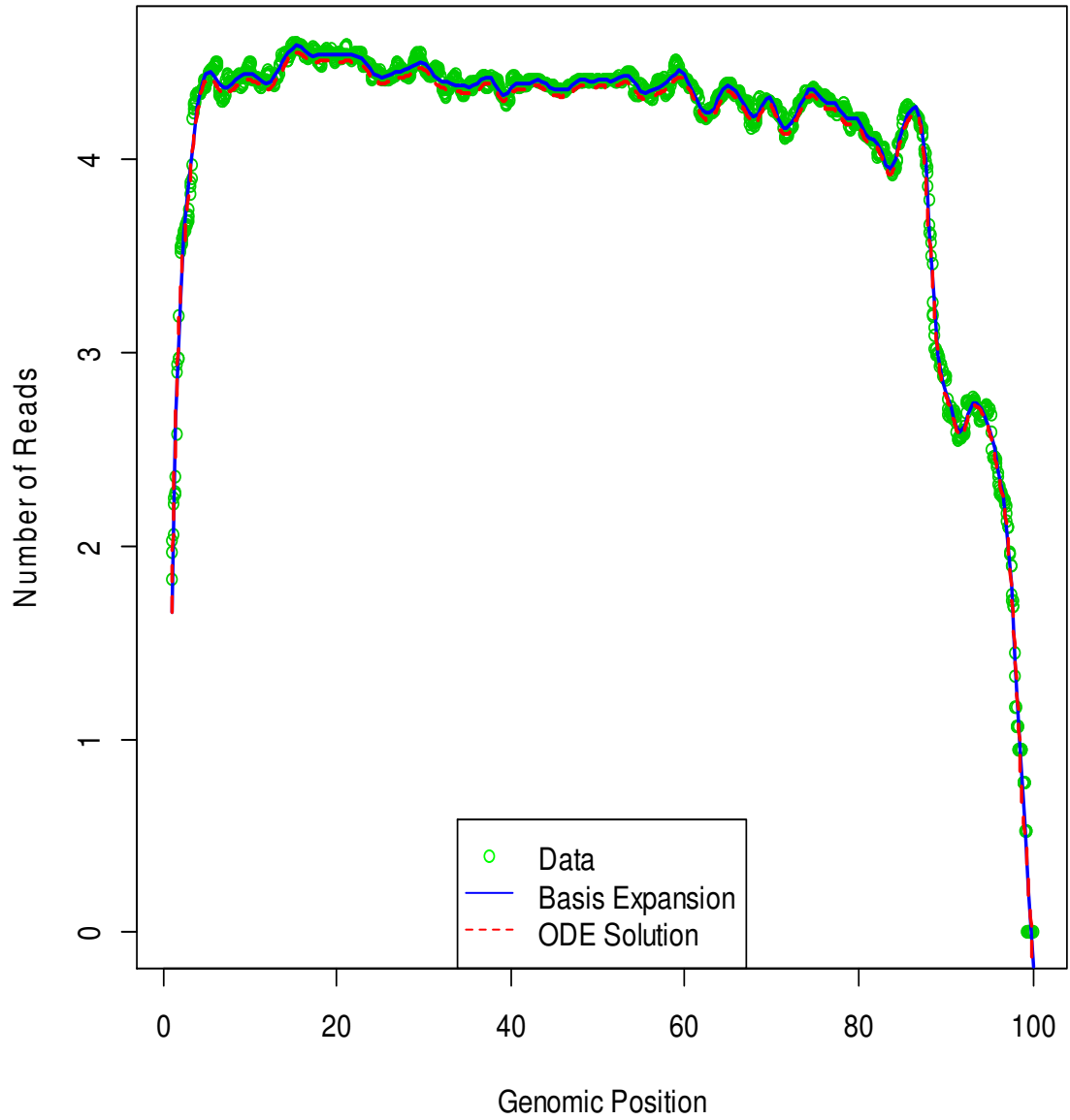

Figure 1 A

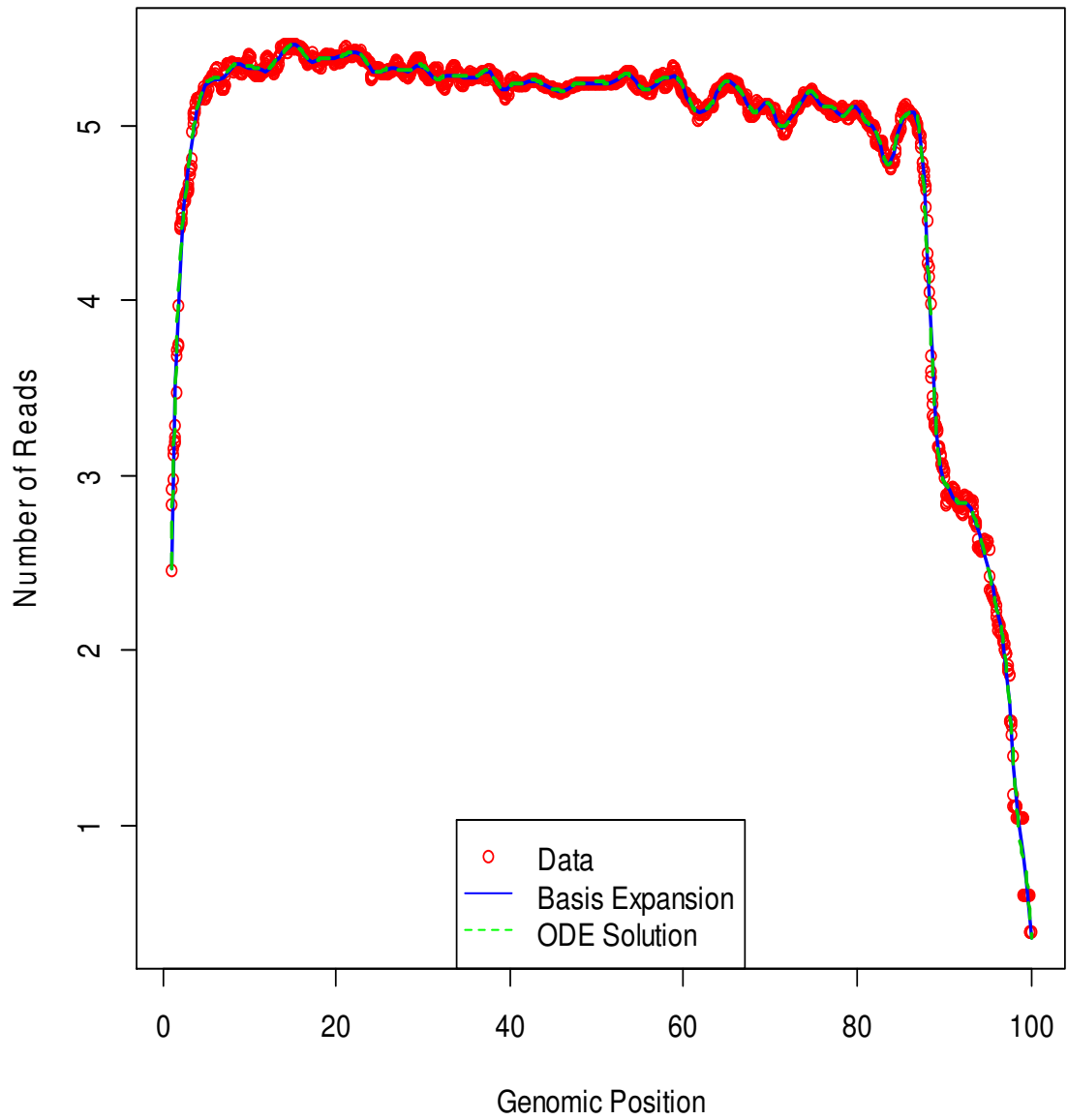

Figure 1B

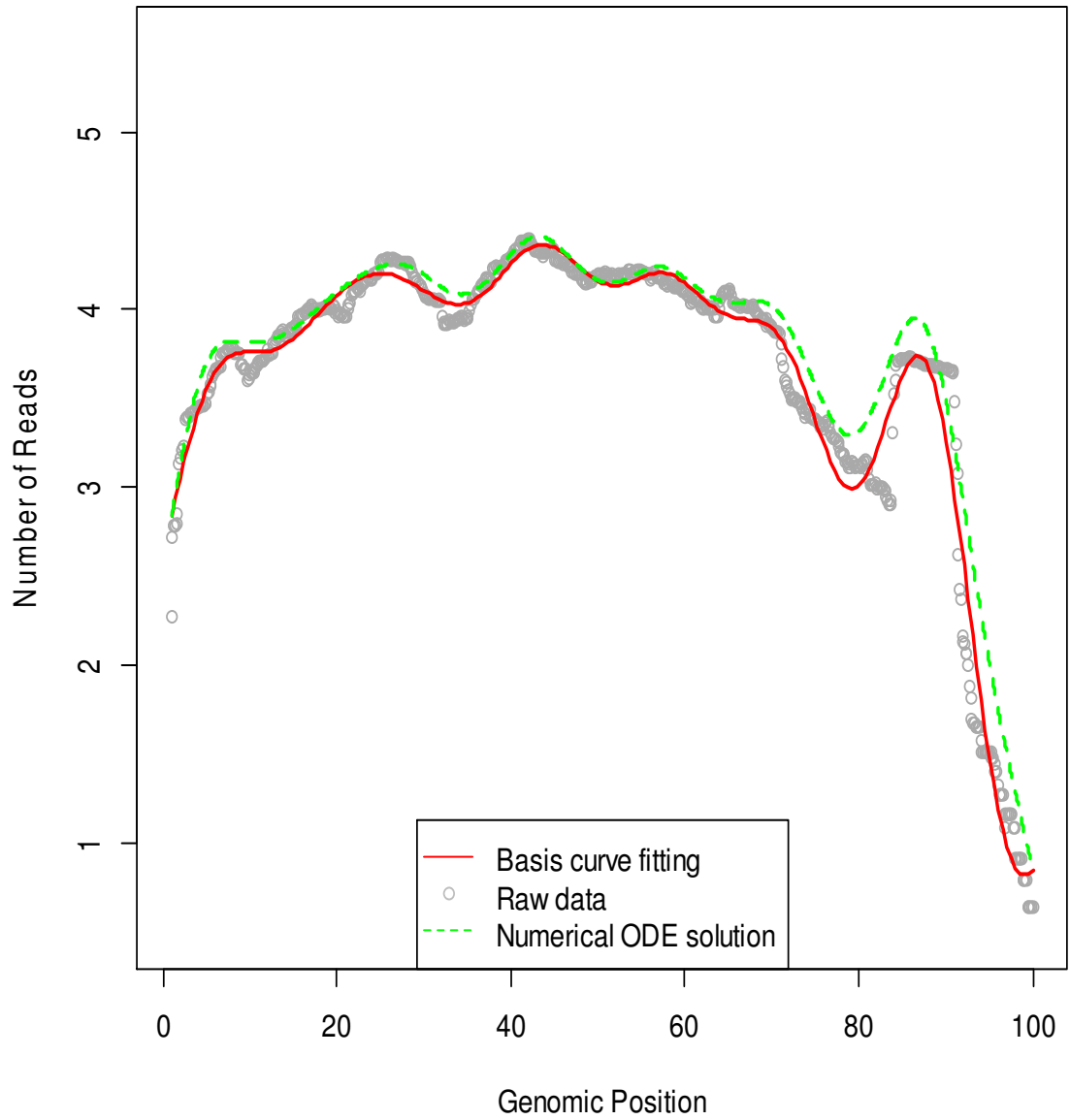

Figure 2A

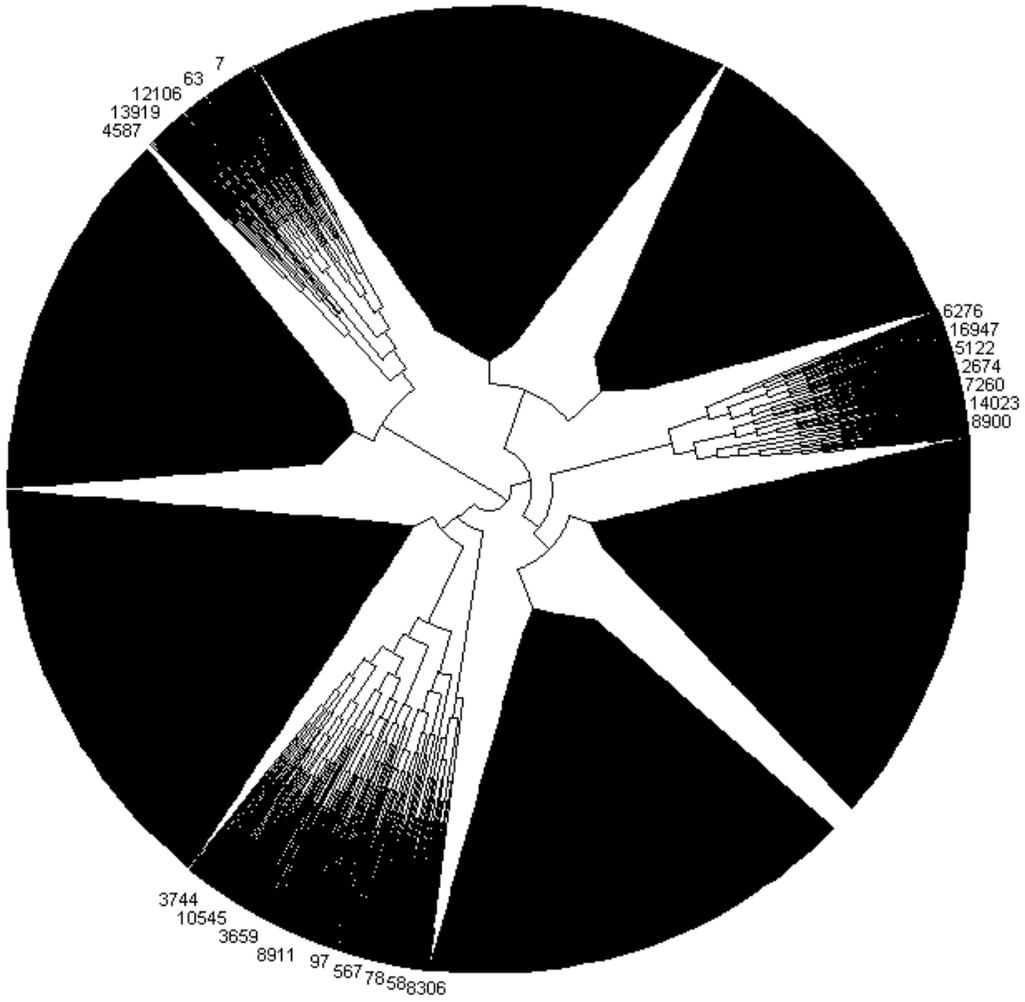

Figure 3A

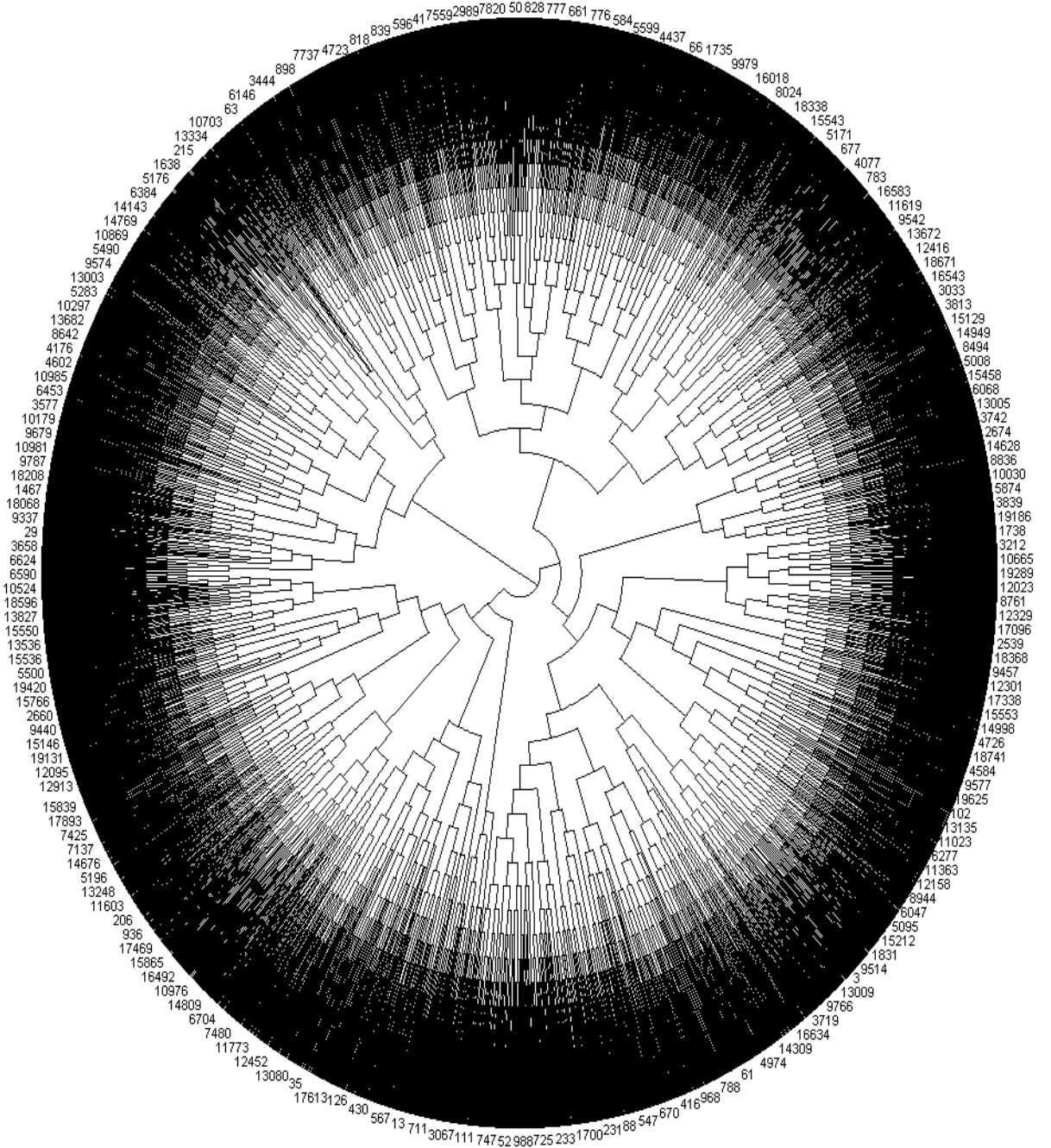

Figure 3B

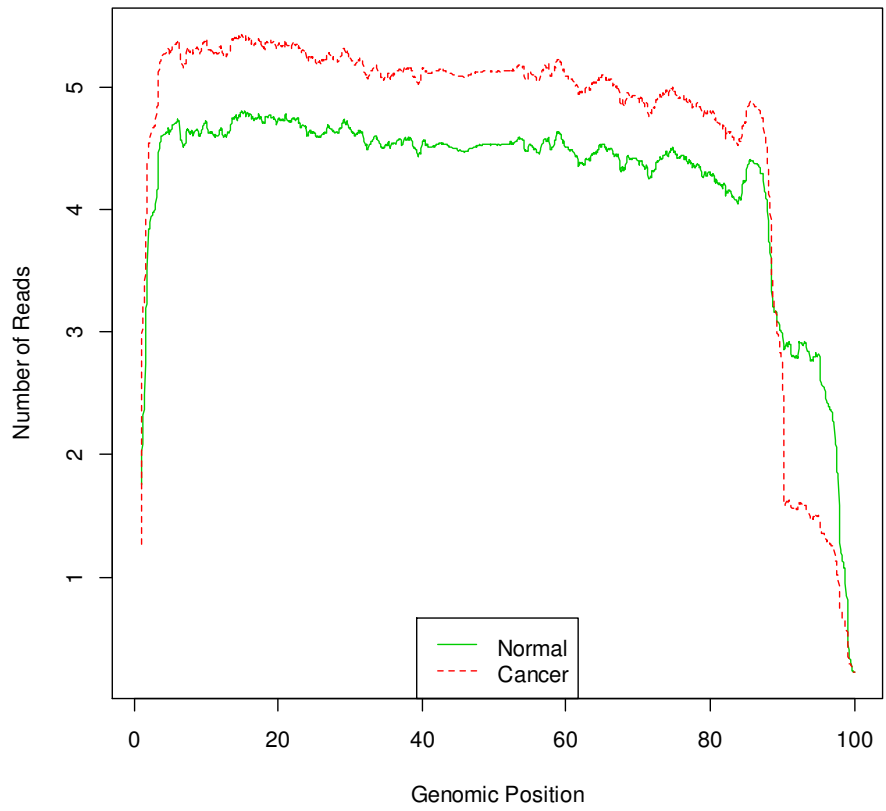

Figure 4A

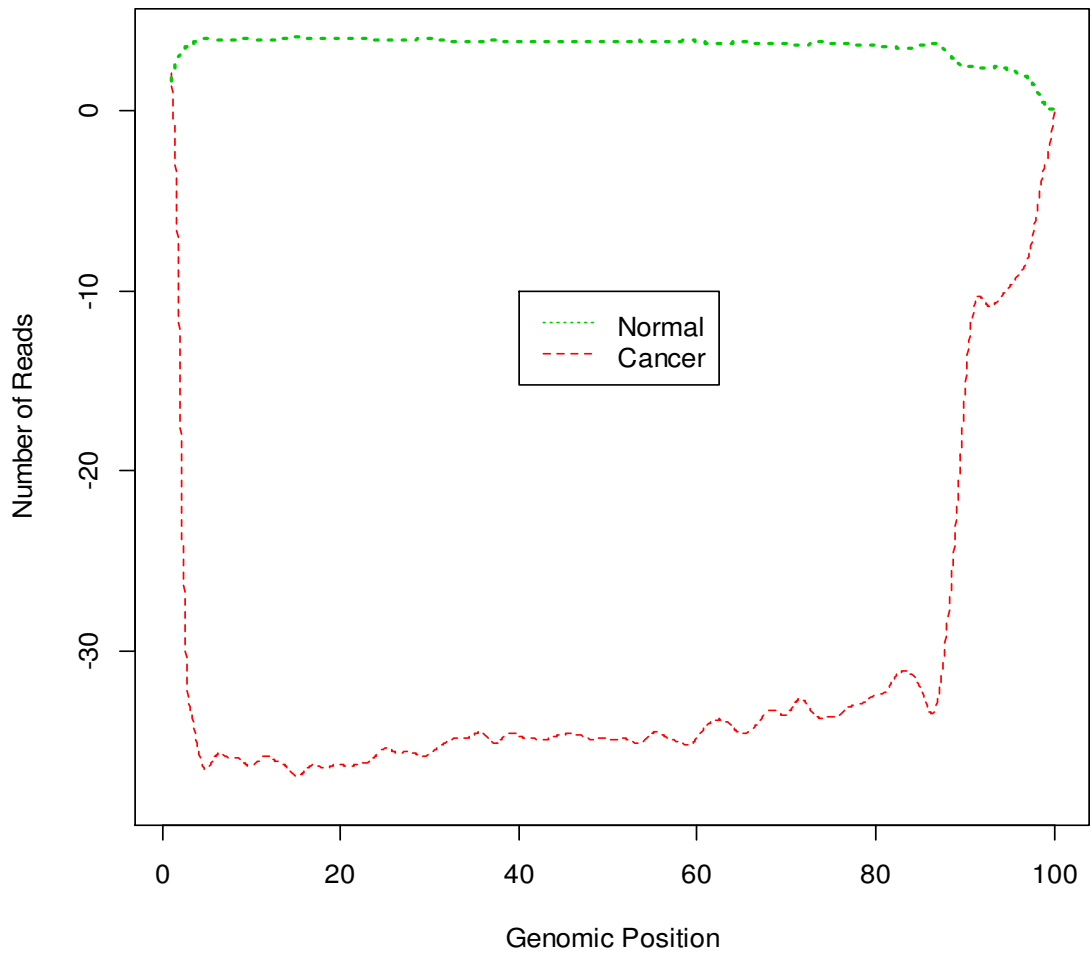

**Figure 4B**

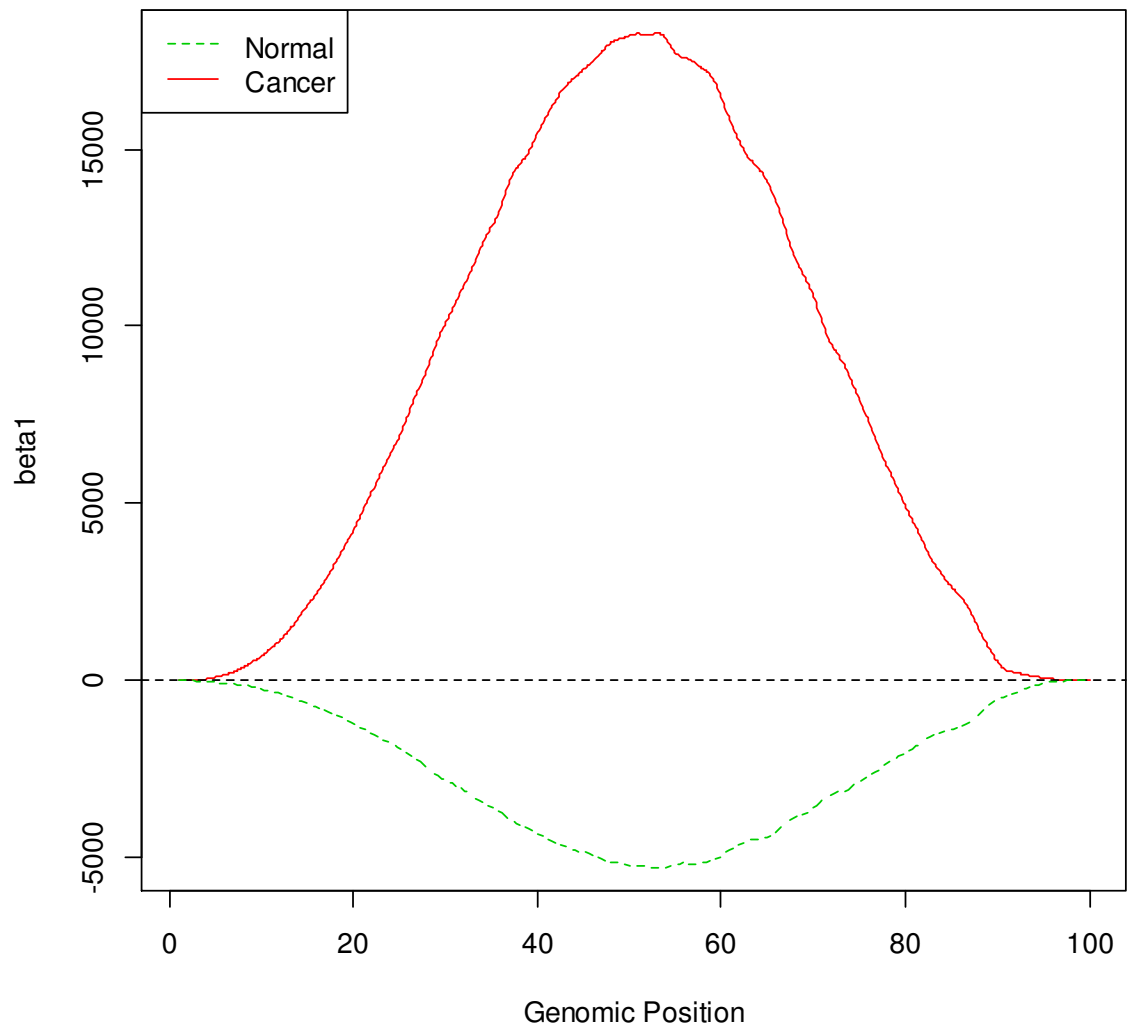

**Figure 4C**

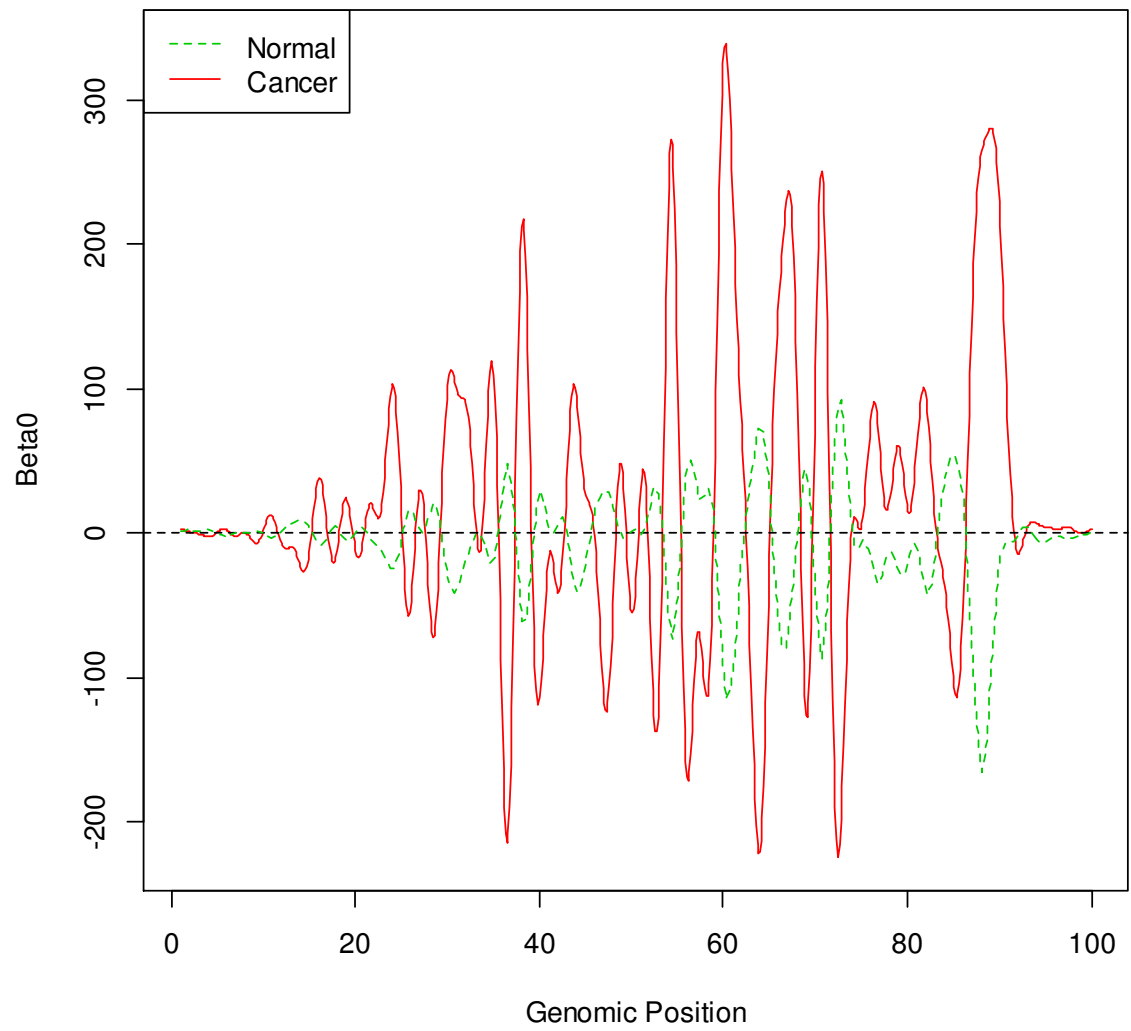

**Figure 4D**

**Figure Legend**

**Figure 1A.** Estimate of expression profiles for CD74 by the ODE in a randomly selected normal sample. The green dotted points were observed expression levels, the blue solid lines are Fourier basis expansions and the red dashed lines are numerical solution of ODE model.

**Figure 1B.** Estimate of expression profiles for CD74 by the ODE in a randomly selected tumor sample. The red dotted points were observed expression levels, the blue solid lines are Fourier basis expansions and the green dashed lines are numerical solution of ODE model.

**Figure 2A.** Predicted expression curves for normal tissues for Gene RPL29: The gray dot is observed expression profile, and the solid red lines are Fourier basis expansion approximation to the observed expression data. The dashed green lines are predicted gene expression profile in the test set by solving the estimated ODE in the training examples.

**Figure 2B.** Predicted expression curves for tumor tissues for Gene RPL29: The gray dot is observed expression profile, and the solid red lines are Fourier basis expansion approximation to the observed expression data. The dashed green lines are predicted gene expression profile in the test set by solving the estimated ODE in the training examples.

**Figure 3A.** Circular phylogram tree of 19717 gene that were clustered into nine gropus by Dendroscope 3.2.10.

**Figure 3B.** Detailed Circular phylogram tree of 19717 gene that were clustered into nine gropus by Dendroscope 3.2.10.

**Figure 4A.** Average expression curves of gene CD74 in normal and tumor samples.

**Figure 4B.** Average unit-step response curves of gene CD74 in normal and tumor samples.

**Figure 4C.** Average coefficient curves of the ODE for gene CD74.

**Figure 4D.** Average coefficient curves of the ODE for gene CD74.